\documentclass[12pt,a4wide]{article}
\usepackage{graphicx}
\usepackage{amsmath}
\usepackage{amssymb}
\usepackage{bm}
\begin{document}

\title{
\bf \large \bf Reduction of static field equation of Faddeev model\\
 to first order PDE\\[30pt]}
\author{Minoru Hirayama$^{1}$ and Chang-Guang Shi $^{2}$\\[10pt]
\\{\normalsize  {\sl Department of Mathematics and Physics, Shanghai University of Electric Power, }}
\\{\normalsize {\sl Pinglian Road 2103, Shanghai 200090, China}}}

\date{}
\maketitle

\begin{abstract}
A method to solve the static field equation of the Faddeev model is presented.
For an special combination of the concerned field, we adopt a form which is compatible with the field equation and involves two arbitrary complex functions.
As a result, the static field equation is reduced to a set of first order partial differential equations.
\end{abstract}

\vspace{3cm}
---------------------------------------------------------------------------------------------

\small{Corresponding author: Chang-Guang Shi,
shichangguang@shiep.edu.cn}

\newpage

Faddeev model\cite{Faddeev} is an effective field theory which is expected to describe the low energy behavior of the $SU(2)$ gauge field\cite{FN}. Its numerical solutions exhibit quite interesting knot-soliton properties\cite{BS,JP}. Because of the high nonlinearity of the model, however, its analytic analysis does not seem to have shown much progress.
The purpose of this letter is to present a method to integrate the static field equation once and obtain a set of first order partial differential equations.\\
   The model concerns the real scalar fields
\begin{equation}
\boldsymbol{n}(x)=\left(n^1(x),n^2(x),n^3(x)\right)
\end{equation}
 satisfying
\begin{equation}
{\boldsymbol{n}}^2(x)=\boldsymbol{n}(x)\cdot\boldsymbol{n}(x)=\sum\limits_{a=1}^{3}n^a(x) n^a(x)=1.
\end{equation}
The Lagrangian density of the Faddeev model is given by
\begin{align}
{\mathcal L}_F(x)&=c_2 l_2(x)+c_4 l_4(x),\\
l_2(x)&=\partial_{\mu}\boldsymbol{n}(x)\cdot
 \partial^{\mu}\boldsymbol{n}(x),\\
l_4(x)&=-H_{\mu\nu}(x) H^{\mu\nu}(x),\\
H_{\mu\nu}(x)&=\boldsymbol{n}(x)\cdot[\partial_{\mu}\boldsymbol{n}(x)\times
 \partial_{\nu}\boldsymbol{n}(x)]\nonumber\\
&=\epsilon_{abc}n^a(x)\partial_{\mu}n^b(x)\partial_{\nu}n^c(x),
\end{align}
where $c_2$ and $c_4$ are constants. The static energy
functional $E_F[\boldsymbol{n}]$ associated with ${\mathcal L}_F(x)$ is
given by
\begin{align}
E_F[\boldsymbol{n}]&=\int dV \epsilon(\boldsymbol{x}),\\
 \epsilon(\boldsymbol{x})&=c_2 \epsilon_2(\boldsymbol{x})+c_4
 \epsilon_4(\boldsymbol{x}),\\
\epsilon_2(\boldsymbol{x})&=\sum\limits_{a=1}^{3}\sum\limits_{i=1}^{3}[\partial_i
 n^a(\boldsymbol{x})]^2,\\
\epsilon_4(\boldsymbol{x})&=\sum\limits_{i,j=1}^{3}[H_{ij}(\boldsymbol{x})]^2,
\end{align}
with $\boldsymbol{x}=(x_1,x_2,x_3)$ and $dV=dx_1 dx_2 dx_3$.
The field $\bm{n}$ can be expressed by a complex function $u$ as
 \begin{equation}
{\boldsymbol n}=\biggl(\frac{u+u^{*}}{|u|^2+1},\frac{-i(u-u^{*})}{{|u|}^2+1},\frac{{|u|}^2-1}{{|u|}^2+1}\biggr).
\end{equation}
In terms of $u$, the energy densities $\epsilon_2$ and $\epsilon_4$ are given by
\begin{equation}
\epsilon_2=\frac{4}{(1+|u|^2)^2}(\boldsymbol{\nabla} u\cdot\boldsymbol{\nabla} u^{*}),
\end{equation}
\begin{align}
\epsilon_4=-8\frac{({\boldsymbol \nabla}u\times{\boldsymbol \nabla}u^*)^2}{(1+|u|^2)^4}.
\end{align}

We here define $R, \Phi, X$ and $\bm{q}$ by
\begin{align}
R&=|u|,\quad u=R{\rm{e}}^{i\Phi},\\
X&=2\sqrt{\frac{c_4}{c_2}}\frac{1}{1+R^2}=\frac{1}{1+R^2},\\
\bm{q}&=X\nabla u,
\end{align}
where $2\sqrt{c_4/c_2}$ of the dimension of length has been set equal to $1$ and $\bm{q}$ is dimensionless.
It can be seen that the derivatives of the static energy density
$\epsilon=c_2 \epsilon_2+c_4 \epsilon_4$
with respect to $u$ and $\nabla u$ are given by
\begin{align}
\frac{\partial\epsilon}{\partial u}&=\frac{-8c_2u^{\star}}{(1+R^2)^3}(\nabla u\cdot \bm{v}),\\
\frac{\partial\epsilon}{\partial (\nabla
u)}&=\frac{4c_2}{(1+R^2)}\bm{v},
\end{align}
respectively, where $\bm{v}$ is defined by
\begin{equation}
\bm{v}=\nabla u^{\star}-X^2\nabla u^{\star}\times(\nabla u\times\nabla u^{\star}).
\end{equation}
Then the static field equation can be written as
\begin{equation}
\nabla\cdot\bm{\alpha}+i \bm{\beta}\cdot\bm{\alpha}=0
\label{eqn:alphaEq}
\end{equation}
and its complex conjugate, where $\bm{\alpha}$ and $\bm{\beta}$ are a complex and a real $3$-vectors given by
\begin{align}
\bm{\alpha}&=\bm{q}^{\star}-\bm{q}^{\star}\times(\bm{q}\times\bm{q}^{\star}),\label{eqn:alphaDEF}\\
\bm{\beta}&=\frac{1}{i}(u^{\star}\bm{q}-u\bm{q}^{\star})= B\nabla
\Phi,~B=\frac{2R^2}{1+R^2},
\end{align}
respectively.
If we define $\bm{\alpha}_i~ (i=1,2)$ by
\begin{align}
\bm{\alpha}_1&={\rm{e}}^{-iB\Phi}(\nabla R\times\nabla \rho),\\
\bm{\alpha}_2&=\nabla \Phi \times \nabla \mu,
\end{align}
we find that they satisfy $\nabla\cdot\bm{\alpha}_i+i \bm{\beta}\cdot\bm{\alpha}_i=0$ for arbitrary differentiable complex functions $\rho$ and $\mu$.
Then, it is clear that $\bm{\alpha}$ defined by
\begin{equation}
\bm{\alpha}=\bm{\alpha}_1+\bm{\alpha}_2
\end{equation}
also satisfies Eq.(\ref{eqn:alphaEq}).
Although it is not clear  if the above form of $\bm{\alpha}$ is general enough or not, it keeps some generality that it involves two arbitrary complex functions. It is interesting that the static field equation of the Faddeev model possesses the above kind of linearity.\\

We hereafter consider how the functions $\rho$ and $\mu$, which are arbitrary at present, are restricted.
We regard $\rho$ and $\mu$ as functions of $R, \Phi$ and $\zeta$ satisfying $\displaystyle{\frac{\partial(R, \Phi,\zeta)}{\partial(x_1,x_2,x_3)}\neq 0}$, where $\zeta$ is a real function obeying
\begin{equation}
\nabla \zeta=\gamma \nabla R\times R\nabla \Phi+\xi \nabla R+R\eta\nabla \Phi.
\end{equation}
Here $\gamma, \xi$ and $\eta$ are real functions. Substituting
$\nabla \rho=\rho_R \nabla R+\rho_{\Phi} \nabla \Phi+\rho_{\zeta}
\nabla \zeta$ with ${\rho}_R=\displaystyle{\frac{\partial
\rho}{\partial R}}$, etc., in $\bm{\alpha}$, we obtain
\begin{align}
 \bm{\alpha}&=\frac{1}{R}\left[{\rm{e}}^{-iB\Phi} \left(\rho_{\Phi}+R\eta\rho_{\zeta}\right)-\left(\mu_{R}+\xi\mu_{\zeta}\right)\right](\bm{y}\times\bm{z})\nonumber\\
&+\frac{\gamma}{R}\left({\rm{e}}^{-iB\Phi} R q \rho_{\zeta}+r \mu_{\zeta}\right)\bm{y}\nonumber\\
&+\frac{\gamma}{R}\left(-{\rm{e}}^{-iB\Phi} R p \rho_{\zeta}-q
\mu_{\zeta}\right)\bm{z}, \label{eqn:alpha1}
  \end{align}
  where $\bm{y}, \bm{z}, p,q$ and $r$ are defined by
  \begin{align}
  \bm{y}=\nabla R,~\bm{z}=R\nabla \Phi,\\
  p=\bm{y}^2,~q=\bm{y}\cdot\bm{z},~r=\bm{z}^2.
   \end{align}
On the other hand, from Eq.(\ref{eqn:alphaDEF}) and $\bm{q}={\rm{e}}^{i\Phi} X(\bm{y}+i\bm{z})$ ,  we have
\begin{equation}
\bm{\alpha}={\rm{e}^{-i\Phi}} \left\{ \left[ (X+rY)+iqY\right] \bm{y}+\left[ -qY-i(X+pY)\right] \bm{z} \right\},
\label{eqn:alpha2}
\end{equation}
where $Y$ is defined by
\begin{equation}
Y=2X^3.
\end{equation}
Irrespectively of the properties of the real $3$-vectors $\bm{y}$ and $\bm{z}$, two expressions (\ref{eqn:alpha1}) and (\ref{eqn:alpha2}) for $\bm{\alpha}$ coincide with each other when $\rho$ and $\mu$ satisfy
\begin{align}
\rho_{\Phi}+R\eta\rho_{\zeta}&= {\rm{e}}^{iB\Phi}\left(\mu_{R}+\xi\mu_{\zeta}\right),\label{eqn:rhomu}\\
\gamma\left({\rm{e}}^{-iB\Phi} R q \rho_{\zeta}+r \mu_{\zeta}\right)&={\rm{e}}^{-i\Phi}R\left[ (X+rY)+iYq\right],\label{eqn:pqr1}\\
\gamma\left({\rm{e}}^{-iB\Phi} R p \rho_{\zeta}+q \mu_{\zeta}
\right)&= {\rm{e}}^{-i\Phi}R\left[
qY+i(X+pY)\right].\label{eqn:pqr2}
\end{align}
Eqs.(\ref{eqn:pqr1}) and (\ref{eqn:pqr2}) constitute four real and linear equations for three real quantities $p, q$ and $r$.
Defining real functions $a,b,c$ and $d$ by
\begin{align}
a&=\gamma {\rm{Re}}\left[{\rm{e}}^{i(1-B)\Phi}\rho_{\zeta}\right],\\
b&=\frac{\gamma}{R}{\rm{Re}}\left[{\rm{e}}^{i\Phi}\mu_{\zeta}\right]-Y,\\
c&=\gamma {\rm{Im}}\left[{\rm{e}}^{i(1-B)\Phi}\rho_{\zeta}\right]-Y,\\
d&=\frac{\gamma}{R}{\rm{Im}}\left[{\rm{e}}^{i\Phi}\mu_{\zeta}\right],
\end{align}
Eqs.(\ref{eqn:pqr1}) and (\ref{eqn:pqr2}) are seen to be equivalent to
\begin{align}
p&=\frac{-bX}{ad-bc},\label{eqn:p}\\
q&=\frac{aX}{ad-bc}=\frac{dX}{ad-bc},\label{eqn:q}\\
r&=\frac{-cX}{ad-bc}.\label{eqn:r}
\end{align}
The above results as well as the definitions of $p,q,r$ lead us to
\begin{equation}
a=d
\end{equation}
and
\begin{equation}
b\geq 0,~c\geq 0,~ad-bc\leq 0.
\label{eqn:ineq}
\end{equation}
Up to now, we have made use of (\ref{eqn:pqr1}), (\ref{eqn:pqr2}) of
the three conditions (\ref{eqn:rhomu}), (\ref{eqn:pqr1}),
(\ref{eqn:pqr2}). \\

Now we discuss how (\ref{eqn:rhomu}) can be used
to obtain further restrictions on $a,b$ and $c$. With making use of
$a=d$ and the definitions of $a,b,c$ and $d$, we have
\begin{align}
\rho_{\zeta}&=\frac{{\rm{e}}^{i(B-1)\Phi}}{\gamma}\left[a+i(c+Y)\right],\label{eqn:rhozeta}\\
\mu_{\zeta}&=\frac{{\rm{e}}^{i(B-1)\Phi}R}{\gamma}\left[(b+Y)+ia\right]\label{eqn:muzeta}.
\end{align}
From (\ref{eqn:rhozeta}), (\ref{eqn:muzeta}) and
\begin{equation}
\left(\rho_{\zeta}\right)_{\Phi}=\left(\rho_{\Phi}\right)_{\zeta}=\left[{\rm{e}}^{iB\Phi}(\mu_R+\xi\mu_\zeta)-R\eta\rho_\zeta\right]_{\zeta},
\end{equation}
or
\begin{equation}
\left(\mu_{\zeta}\right)_{R}=\left(\mu_{R}\right)_{\zeta}=\left[{\rm{e}}^{-iB\Phi}(\rho_{\Phi}+R\eta\rho_\zeta)-\xi\mu_\zeta\right]_{\zeta},
\end{equation}
we obtain two real relations:
\begin{align}
\left[\frac{R(b+Y)}{\gamma}\right]_R-\left(\frac{a}{\gamma}\right)_{\Phi}&=\frac{(1-B)(c+Y)}{\gamma}+\left\{\frac{R\left[\eta
a-\xi(b+Y)\right]}{\gamma}\right\}_{\zeta},\\
\left(\frac{Ra}{\gamma}\right)_R-\left(\frac{c+Y}{\gamma}\right)_{\Phi}
&=\frac{(B-1)a}{\gamma}+\left\{\frac{R\left[\eta (c+Y)-\xi
a\right]}{\gamma}\right\}_{\zeta}.
\end{align}
These relations assure the existence of the complex functions $\rho$ and $\mu$ satisfying (\ref{eqn:rhomu}).
Defining differential operators $d_1, d_2, D_1$ and $D_2$ by
\begin{align}
d_1&=\frac{\partial}{\partial R}+\xi\frac{\partial}{\partial \zeta},\\
d_2&=\frac{\partial}{\partial \Phi}+R\eta\frac{\partial}{\partial \zeta},\\
D_1&= R d_1+1+R \xi_{\zeta},\\
D_2&=d_2+R\eta_{\zeta},
\end{align}
they are rewritten as
\begin{align}
\begin{pmatrix}
d_1(\ln \gamma)\\
d_2(\ln \gamma)
\end{pmatrix}
=\frac{1}{G}
\begin{pmatrix}
-(c+Y)& a\\
-Ra &R(b+Y)
\end{pmatrix}
\begin{pmatrix}
E\\
F
\end{pmatrix}\label{eqn:lngamma}
\end{align}
with
\begin{align}
&E=(B-1)(c+Y)+D_1(b+Y)-D_2(a),\\
&F=(1-B)a+D_1(a)-D_2((c+Y),\\
&G=R[a^2-(b+Y)(c+Y)].
\end{align}
We have thus found the relation that $\gamma, \xi, \eta, a, b$ and $c$ should satisfy. It consists of two partial differential equations of first order.\\

To get more insight for the relation (\ref{eqn:lngamma}), we here consider a special case in which $\gamma, \xi$ and $\eta$ are fixed independently of $a, b$ and $c$. If we assume that $\xi$ and $\eta$ satisfy
\begin{equation}
d_1(R\eta)-d_2(\xi)=\eta+R\eta_R+R\left(\xi\eta_{\zeta}-\eta\xi_{\zeta}\right)-\xi_{\Phi}=0,
\label{eqn:etazeta}
\end{equation}
then we have
\begin{align}
\left[ d_1, d_2 \right]=0,\quad \quad \left[ D_1, D_2 \right]=0
\end{align}
and we are allowed to choose $\gamma$ so as to satisfy
\begin{align}
\begin{pmatrix}
d_1(\ln \gamma)\\
d_2(\ln \gamma)
\end{pmatrix}
=\kappa
\begin{pmatrix}
\xi\\
R\eta
\end{pmatrix}.
\label{eqn:gamma}
\end{align}
If we further assume that  $\xi$ is a function of $\zeta$ multiplied by a function of $R$ and $\Phi$, Eqs.(\ref{eqn:etazeta}), (\ref{eqn:gamma}) lead us to the conclusion
\begin{align}
\xi=\frac{1}{k'(\zeta)}&\frac{\partial j(R, \Phi)}{\partial R},\quad k'(\zeta)=\frac{dk(\zeta)}{d\zeta},\\
R\eta&=\frac{1}{k'(\zeta)}\frac{\partial j(R, \Phi)}{\partial \Phi},\\
\gamma&={\rm{e}}^{\kappa \zeta} w\left(j(R, \Phi)-k(\zeta)\right),
\end{align}
where $k, j$ and $w$ are arbitrary real functions. If we substitute these results in (\ref{eqn:lngamma}), we obtain two relations on $a,b$ and $c$.
When we specify $k(\zeta)$ as
\begin{equation}
k'(\zeta)={\rm{e}}^{-\kappa\zeta},
\label{eqn:kappa}
\end{equation}
we obtain rather simple relations for $(a, b, c)$:
\begin{align}
R d_1(b+Y)-d_2(a)&=-(b+Y)+(1-B)(c+Y),\\
R d_1(a)-d_2(c+Y)&=(B-2)a.
\end{align}
The specification (\ref{eqn:kappa}) also yields the expressions
\begin{align}
d_1&={\rm{e}}^A\frac{\partial}{\partial R}{\rm{e}}^{-A},~d_2={\rm{e}}^A\frac{\partial}{\partial \Phi}{\rm{e}}^{-A},\\
A&=-{\rm{e}}^{\kappa\zeta}j(R, \Phi)\frac{\partial}{\partial\zeta}
\end{align}
for $d_1$ and $d_2$.
Defining $L, M$ and $N$ by
\begin{equation}
L={\rm{e}}^{-A}(b+Y),~M={\rm{e}}^{-A}a,~N={\rm{e}}^{-A}(c+Y),
\end{equation}
we arrive at a simple set of equations
\begin{align}
R \frac{\partial L}{\partial R}-\frac{\partial M}{\partial \zeta}&=-L+(1-B)N,\\
R \frac{\partial M}{\partial R}-\frac{\partial
N}{\partial\Phi}&=(B-2)M.
\end{align}
It turns out that the general solution of (70) and (71) involves five arbitrary functions in addition to $j(R, \Phi)$. It seems that the solution $(a, b, c)$ possesses enough flexibility to attain the inequalities (\ref{eqn:ineq}).   More thorough analysis will be given elsewhere.\\

After we find $a,b,c, \gamma, \xi$ and $\eta$ satisfying the relation (\ref{eqn:lngamma}), we are left with the first order partial differential equations of the following form:
\begin{align}
&p=(\nabla R)^2=S(R,\Phi,\zeta),\\
&q=\nabla R\cdot R\nabla \Phi=T(R,\Phi,\zeta),\\
&r=(R\nabla \Phi)^2= U(R,\Phi,\zeta),
\end{align}
where $S, T$ and $U$ are the functions fixed by (\ref{eqn:p}), (\ref{eqn:q}), (\ref{eqn:r}).
When we solve the last equations, the topological consideration concerning the Hopf charge which classifies the mapping $S^3\rightarrow S^2$ would be necessary and, in such a situation, the variable $\zeta$ might become important.
Since $X^2(p+r)$ and $X^4(pr-q^2)$ are proportional to $\epsilon_2$ and $\epsilon_4$, respectively, the finiteness of $E_F[\bm{n}]$ is also governed by the above $S, T$ and $U$. In conclusion, guided by (25), we have reduced the field equation of the Faddeev model to first order partial differential equations for $R$ and $\Phi$. 

\section*{Acknowledgments}
One of the authors(M. H.) is  grateful to Jun Yamashita for discussions at the early stage of this research. This research was partially supported by the National Natural Science Foundation of China (Grant No. 10601031) and Science Foundation of Shanghai Municipal  Education Commission of China under Grant No.05LZ08.

\end{document}